\documentclass[pra,twocolumn,showpacs]{revtex4-2}

\usepackage{amsmath}
\usepackage{latexsym}
\usepackage{amssymb}
\usepackage{graphicx}
\usepackage[colorlinks=true, citecolor=blue, urlcolor=blue]{hyperref}
\usepackage{float}
\usepackage{amsfonts}
\usepackage{textcomp}
\usepackage{mathpazo}
\usepackage{comment}
\usepackage{xr}

\usepackage{bbm}

\usepackage{xcolor}
\definecolor{myurlcolor}{rgb}{0,.5,.5}
\definecolor{mycitecolor}{rgb}{0,.6,0}
\definecolor{myrefcolor}{rgb}{2,0,0}
\usepackage{hyperref}
\hypersetup{colorlinks,
linkcolor=myrefcolor,
citecolor=mycitecolor,
urlcolor=myurlcolor}

\usepackage[draft]{fixme}
\usepackage{empheq}
\usepackage{amsmath,bbm}
\usepackage{graphicx}
\usepackage{amsfonts}
\usepackage{amssymb}
\usepackage{amsmath, amssymb, amsthm,verbatim,graphicx,bbm}
\usepackage{mathrsfs}
\usepackage{color,xcolor,longtable}

\usepackage{xr}
\makeatletter
\newcommand*{\addFileDependency}[1]{
  \typeout{(#1)}
  \@addtofilelist{#1}
  \IfFileExists{#1}{}{\typeout{No file #1.}}
}
\makeatother

\newcommand*{\myexternaldocument}[1]{
    \externaldocument{#1}
    \addFileDependency{#1.tex}
    \addFileDependency{#1.aux}
}

\myexternaldocument{manuscript_PRL}

\newcommand{\beq}[0]{\begin{equation}}
\newcommand{\eeq}[0]{\end{equation}}

\newcommand{\one}{\leavevmode\hbox{\small1\normalsize\kern-.33em1}}

\def\be{\begin{equation}}
\def\ee{\end{equation}}
\def\ben{\begin{eqnarray}}
\def\een{\end{eqnarray}}
\def\eea{\end{array}}
\def\bea{\begin{array}}

\newcommand{\Tr}[1]{\mathrm{Tr}#1}
\newcommand{\bei}{\begin{itemize}}
\newcommand{\eei}{\end{itemize}}
\newcommand{\ket}[1]{|#1\rangle}
\newcommand{\bra}[1]{\langle#1|}

\newcommand{\proj}[1]{\ket{#1}\!\!\bra{#1}}

\newcommand{\I}{\mathbbm{1}}

\newcommand{\dl}[1]{\left|\left|#1\right|\right|}

\renewcommand{\emph}[1]{\textbf{#1}}

\makeatletter
\newtheorem*{rep@theorem}{\rep@title}
\newcommand{\newreptheorem}[2]{%
\newenvironment{rep#1}[1]{%
 \def\rep@title{#2 \ref{##1}}%
 \begin{rep@theorem}}%
 {\end{rep@theorem}}}
\makeatother

\theoremstyle{plain}
\newtheorem*{thm*}{Theorem}
\newreptheorem{thm}{Theorem}

\newtheorem*{ucor}{Corollary}

\theoremstyle{definition}

\theoremstyle{remark}

\usepackage[T1]{fontenc}

\begin{document}

\title{Distrustful entanglement-based Quantum Key Distribution}
\author{Shubhayan Sarkar}
\email{shubhayan.sarkar@jcbose.ac.in}
\affiliation{Department of Physical Sciences, Bose Institute, EN 80, Bidhannagar, Kolkata 700091, India}

\begin{abstract}
Quantum key distribution (QKD) is the one of the most explored application of quantum information theory. While practical QKD systems are being deployed around the world, a major issue of these setups are that one has to trust the measurement devices of both parties. On the other hand, the most secure among such schemes are called device-independent (DI) QKD protocols, where no component of the setup is assumed to be trusted. However, due to their high technological demand, their loophole-free version currently remain out of practical reach. Inspired by the real world, we consider the traditional QKD in a distrustful measurement setting, that is, the measurement devices can only be trusted upto some level of confidence, which might still be large but not completely trusted. We then derive a general lower bound on the asymptotic key rate in this scenario. Then considering perfect detectors and source, and the entanglement-based BB84 (BBM) protocol, we show that to obtain secure keys one needs to trust their devices more than 92 \%. However, we observe that such a protocol can not be realised with any level of distrust using the realistic detectors and sources at the current technology. Finally, we also consider the BBM protocol with arbitrary dimensional systems and observe that they are more robust to distrust than qubit systems. Consequently, with the same level of distrust, along with same detector and source efficiences, one can get better key rates with high-dimensional systems.  
\end{abstract}

\maketitle

{\it{Introduction---}} With the advent of quantum computers, conventional public-key cryptographic systems, such as RSA, are threatened. To address this threat, institutions throughout the world are shifting towards Quantum Key Distribution (QKD) protocols \cite{QKDrev1,QKDrev2}.
QKD enables the establishment of shared secret keys between spatially separated users with information-theoretic security rooted in the laws of quantum mechanics rather than computational assumptions. The first QKD protocol, BB84 \cite{bennett1984quantum}, encodes information in non-orthogonal quantum states, exploiting the fact that attempts to acquire information about these states inevitably disturb them, enabling the legitimate users to detect eavesdropping. This principle was subsequently extended to entanglement-based protocols, including E91 \cite{Ekert1991E91}, which connects key security to the violation of Bell inequalities, and BBM92 \cite{bennett1992quantum}, the entanglement-based counterpart of BB84. As a matter of fact, QKD is being deployed and tested in fibre-optic and satellite-based communication networks as a potential means of securing sensitive information against future quantum-enabled attacks \cite{expQKD1,expQKD2,expQKD3,QKDep1,QKDep2,QKDep3,QKDep4,QKDep5,QKDep6,QKDep8,QKDep9,QKDep11}, with most of them based on the BB84 (with or without decoy \cite{decoy1, decoy2}) or BBM92 protocol.

The security of all QKD protocols relies on a very strong assumption that the measurement devices are completely trusted, that is, the actual measurement elements are the same as the ones expected such that no attacker can manipulate them. Even a basic item such as a condom is only reliable up to $98\%$, then how can we expect that detectors, which involve complex electronics, are fully reliable? This opens up the possibility of several attacks by an adversary, as was demonstrated first in \cite{attack1}, and subsequently in many other works \cite{attack2, attack3,attack4,attack5,attack6,attack7,attack11,attack8,attack9,attack10}.
Inspired by such problems, the idea of device-independent(DI) QKD (no device trusted) \cite{diqkd2,diqkd3,diqkd5, DIQKDrev1,DIQKDrev2,DIQKDrev3} and one-sided(1S) DIQKD (one measurement device trusted) \cite{1SDIQKD1, 1SDIQKD2, 1SDIQKD3, 1SDIQKD4, 1SDIQKD5} was introduced. However, at the current technology, detector efficiencies are not sufficient for a loophole-free demonstration of DIQKD. While such a loophole-free demonstration is possible for 1SDIQKD protocols, the trust on one side makes them vulnerable to the same attacks as standard QKD protocols.

In recent years, the prepare-and-measure BB84-type device-dependent protocol has been extended to the case when sources and detectors are imperfect \cite{imperfectQKD1, imperfectQKD2, imperfectQKD3, imperfectQKD4, imperfectQKD5, imperfectQKD6,imperfectQKD7,imperfectQKD8,imperfectQKD9,imperfectQKD10,imperfectQKD11,imperfectQKD12,imperfectQKD13,imperfectQKD14,imperfectQKD15,imperfectQKD16,imperfectqkd21,imperfectqkd22}. However, most of the above works treat the source operationally while the detectors are treated based on some physical model of the devices, thus, not operational. While \cite{imperfectqkd21,imperfectqkd22} provide a partial operational approach, however, they restrict the adversary to implement a particular form of the measurement, and thus not completely operational. Moreover, in all the above works the derived key-rate is numerical and thus not general enough to compute key-rates in arbitrary prepare-and-measure scenarios. 

In this work, we introduce the idea of Distrustful entanglement-based Quantum Key Distribution (DEQKD). Here, like the E91 or BBM protocol the source is completely untrusted, while the detectors are partially reliable up to some value of trust. Within this restriction, the adversary can replace the ideal measurement with any other, with no modelling of the measurements involved, similar to DIQKD. Then, for any DEQKD protocol with one-way classical reconciliation, that is, only one party communicates their observed results, we derive a general formula for the lower bound of the key-rate for all such protocols. Interestingly, the formula turns out to be analytically computable and can be expressed simply as the key-rate when the devices are completely trusted, with a correction term depending on the ideal measurements and the trust in the devices. Finally, considering the BBM protocol (generalised for arbitrary dimensions) in this distrustful scenario, we compute the key-rate for various detection efficiencies, source visibilities, levels of trust, and dimensions. Strikingly, we observe that doing practical QKD using qubit systems is only possible with current technology if one fully trusts their devices. Consequently, the vulnerability of currently deployed QKD systems cannot be mitigated unless better detectors are available.

Let us now first describe the entanglement-based QKD protocols. As the security of most QKD protocols that we are aware of is always analysed by considering the equivalent entanglement-based version of the protocol \cite{ShorPreskill2000}, our result is also valid for equivalent prepare-and-measure BB84 type protocols with untrusted sources and measurements.

{\it{Entanglement-based QKD}}---
Consider two parties, namely Alice $(A)$ and Bob $(B)$, who want to establish a secure key among themselves. Both parties receive subsystems from a quantum source, on which they perform at least two measurements each, which are privately chosen, such that the key is extracted from their respective outcomes. The inputs are denoted by $x,y$, while the outputs are denoted by $a,b$. 
The quantum state generated by the source is denoted as $\rho_{AB}$, while the measurements are given by $\{M_{a|x}\},\{M_{b|y}\}$. From the setup, Alice and Bob obtain the correlation given by $\vec{p}=\{p(a,b|x,y)\}$ where each probability is computed using the Born rule.

In a QKD protocol, the rounds are randomly assigned for parameter estimation or key generation. In the parameter estimation rounds, the raw data is publicly disclosed and used to estimate the correlations needed for security proofs. These rounds provide an operational test for the QKD protocol; that is, Alice and Bob compare the observed correlations to the expected ones, and if they do not match, the protocol is aborted. If the test passes, then, the raw key is extracted from the key-generating rounds, which is again post-processed, such as error correction and privacy amplification, to obtain the final key. Here, the classical communication happens in one direction (either $A\rightarrow B$ or $B\rightarrow A$) and thus, these protocols are called QKD protocols with one-way classical reconciliation and are the most studied ones in QKD literature [for instance see Refs. \cite{Ekert1991E91, Bennett1992EntBB84, Tomamichel_2012,  Tomamichel_2017,  QKDrev1, QKDrev2}] and are currently deployed in most practical settings \cite{expQKD1,expQKD2,expQKD3,QKDep1,QKDep2,QKDep3,QKDep4,QKDep5,QKDep6,QKDep8,QKDep9,QKDep11}. 

Let us notice here that the measurement devices are trusted, that is, they are assumed to be implementing the same POVM elements as expected and thus can not be attacked by an eavesdropper. However, no physical detector can always be accurate and can detect every subsystem. The measurements being trusted allows Alice and Bob to neglect the events when the subsystems are undetected, also called the fair-sampling assumption \cite{fair-sampling}.

Consider now that one can trust the measurement devices in the QKD setup upto some parameter $1-\tau$, where we call $\tau$ the distrust parameter. Consequently, we can formulate the above statement for all $a,x$ as
\begin{equation}\label{eq1}
    \dl{M_{a|x}-\tilde{M}_{a|x}}_2\leq \tau
\end{equation}
where $\tilde{M}_{a|x}$ is the ideal measurement POVM while $M_{a|x}$ is the actual measurement POVM in the setup. 
Recall that the source in this setup is anyway untrusted. One can define a similar bound for Bob's measurement elements. We call such protocols as distrustful entanglement-based QKD protocols. Notice that unlike standard QKD protocols, we can not invoke the fair-sampling assumption here. Let us now proceed towards the main result of this work.

\begin{figure*}[t!]
    \centering
    \includegraphics[width=\linewidth]{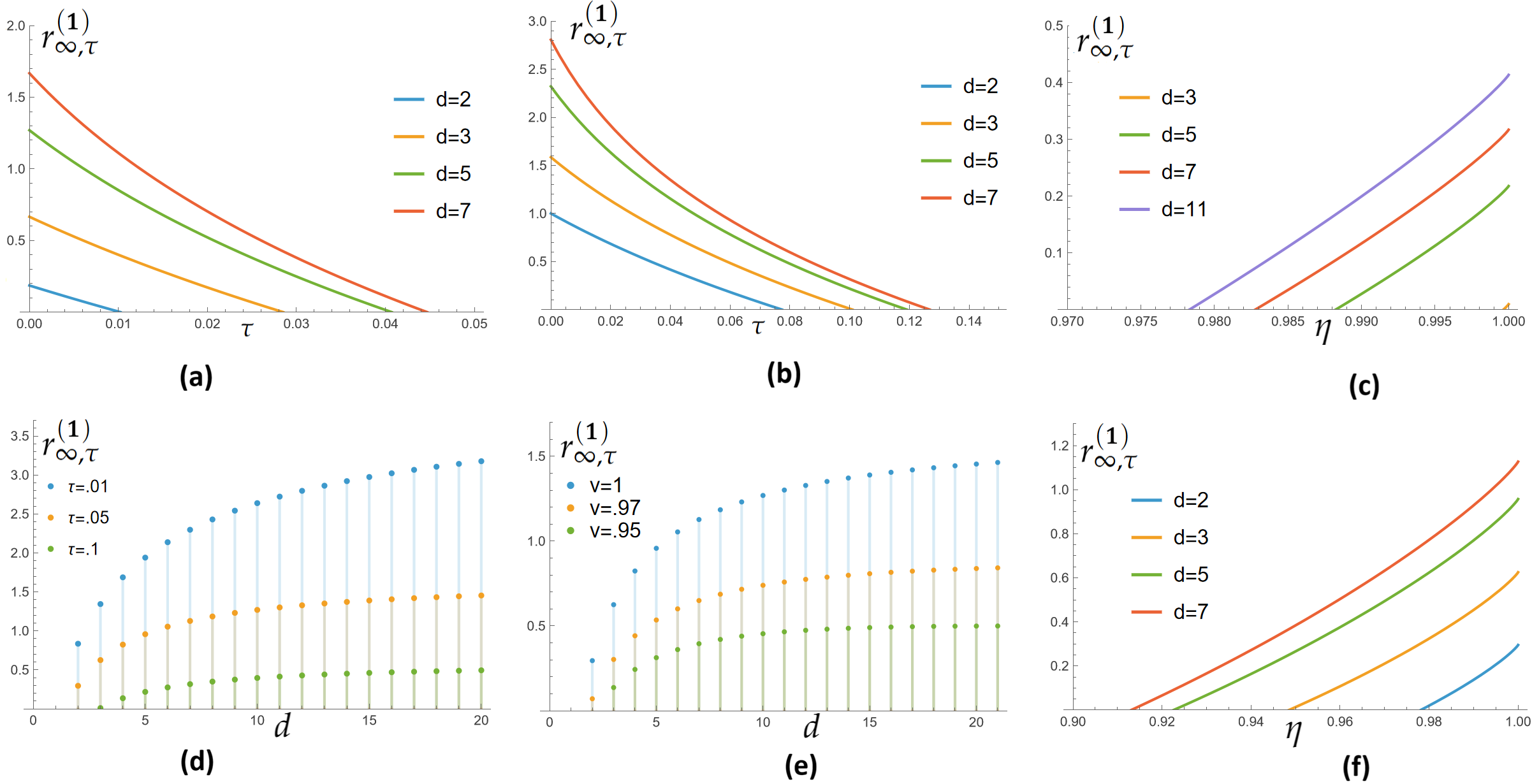}
    \caption {Key rate of the distrustful-generalised-BBM protocol for various parameters. (a) Key rate vs distrust when the detectors are 90 \% efficient, while the source is perfect. (b) Key rate vs distrust when the detectors and the source are perfect. (c) Key rate vs detection efficiency with 90 \% trust for different dimensions when the source is perfect. (d) Key rate vs dimension when the detectors and source are perfect with different levels of trust. (e) Key rate vs dimension with 95 \% trust for different visibilities for 95 \% detection efficiency. (f) Key rate vs detection efficiency with 95 \% trust for different dimensions when the source is perfect.}
    \label{fig1}
\end{figure*}

{\it{Lower-bound on key-rate in distrustful QKD protocols---}} Let us recall that for the above-considered QKD protocols, the lower bound on the asymptotic key-rate for security against collective attacks is given by the Devetak-Winter rate \cite{DevetakWinter2005}
    \begin{equation}\label{rate}
         r_{\infty,\tau}^{(x)}\geq H(A_x|E)-H(A_x|B)
    \end{equation}
    where $r_{\infty,\tau}^{(x)}$ denotes the rate when the key is extracted from the input $x$ of Alice. 

    As we can observe from the key-rate formula \eqref{rate}, while the conditional entropy $H(A_x|B)$ can be upper-bounded using the observed correlations between Alice and Bob, lower-bound to $H(A_x|E)$ has to be computed. For this purpose, let us now consider the entropic uncertainty relation for arbitrary measurements among three parties $A,B,E$ introduced in \cite{Coles_2014}, given as
\begin{equation}\label{EUR1}
    H(A_{x'}|B)+H(A_{x}|E)\geq -\log \Big( \max_{a,a'}\Tr(M_{a|x}M_{a'|x'})\Big),
\end{equation}
From here on, we express the maximal overlap term between Alice's measurement elements as $\mathbbm{O}\equiv\max_{a,a'}\Tr(M_{a|x}M_{a'|x'})$.
Consequently, we obtain from \eqref{rate} that
\begin{eqnarray}\label{mainrate}
    r_{\infty,\tau}^{(x)}\geq -\log \mathbbm{O}-H(A_x|B)- H(A_{x'}|B).
\end{eqnarray}

Let us now compute the maximal overlap between the measurement elements of Alice. 

\begin{thm*}\label{thm1}
    Consider the pair of measurements corresponding to inputs $x,x'$. The maximum overlap $\mathbbm{O}$ of Alice's measurement elements is upper-bounded by 
    \begin{equation}\label{maineq1}
         \mathbbm{O}\leq\max_{a,a'}\Tr(\tilde{M}_{a|x}\tilde{M}_{a'|x'})(1+\Delta(\{\tilde{M}_{a|x}\},\{\tilde{M}_{a'|x'}\},\tau)).
    \end{equation}
where the correction $\Delta(\{\tilde{M}_{a|x}\},\{\tilde{M}_{a'|x'}\},\tau)$ term due to distrust is given by
\begin{eqnarray}\label{correction}
\frac{3\tau\left(\max_{a}\dl{\tilde{M}_{a|x}}+\max_{a'}\dl{\tilde{M}_{a'|x'}}\right)+6\tau^2}{\max_{a,a'}\Tr(\tilde{M}_{a|x}\tilde{M}_{a'|x'})}
\end{eqnarray}
\end{thm*}

The proof of the above theorem is given in Appendix A. If one further assumes that the measurements of Alice are rank-one projective, the above bounds can be further simplified.

\begin{ucor}
    
    Consider the pair of rank-one projective measurements corresponding to inputs $x,x'$. The maximum overlap of the measurement elements is given by 
    \begin{eqnarray}\label{corr1}
        \max_{a,a'}\Tr(M_{a|x}M_{a'|x'})\leq 
         \max_{a,a'}\Tr(\tilde{M}_{a|x}\tilde{M}_{a'|x'})+6\tau+6\tau^2.
    \end{eqnarray}
\end{ucor}
One can also arrive at the above formula \eqref{corr1}, using the entropic-uncertainty relation introduced in \cite{Berta2010}.
Let us now find the rate of key generation for QKD protocols that utilise rank-one projective measurements. 

Consequently, from Theorem \ref{thm1}, we find a general lower bound on the key rate of distrustful QKD protocols with classical one-way reconciliation as
\begin{eqnarray}
 r_{\infty,\tau}^{(x)}\geq   r_{\infty,\tau=0}^{(x)}-\log\left(1+\Delta(\{\tilde{M}_{a|x}\},\{\tilde{M}_{a'|x'}\},\tau)\right)
\end{eqnarray}
such that $r_{\infty,\tau=0}^{(x)}$ is the lower bound to the key-rate when the measurement devices are completely trusted 
\begin{equation}\label{ratetrust}
     r_{\infty,\tau=0}^{(x)}= -\log\left(\max_{a,a'}\Tr(\tilde{M}_{a|x}\tilde{M}_{a'|x'})\right)-\sum_{x,x'}H(A_x|B).
\end{equation}

The above results can straightaway be generalised to the case when multiple bases of Alice are used to extract the key.
Interestingly, one can observe from the above analysis that either Alice's or Bob's distrust is enough to quantify the key rate in this scenario, specifically from the device from which the key is extracted. Consequently, our framework is straightaway applicable to one-sided device-independent QKD protocols \cite{1SDIQKD1,1SDIQKD2,1SDIQKD3,1SDIQKD4,1SDIQKD5} in the distrustful case \cite{disquantsteer1, disquantsteer2}. Let us now analyse the standard entanglement-based BB84 protocol, BBM protocol in arbitrary dimension when the measurements are not completely trusted.

{\it{Distrustful-generalised-BBM-protocol----}}
In the ideal realisation of the BBM protocol, Alice and Bob share the maximally entangled two-qubit state while both perform $\{X,Z\}$ measurements on their respective subsystems, where $Z$ is the computational basis, while $X$ represents the Fourier basis. Here, we consider its generalisation to arbitrary dimension such that the state shared between Alice and Bob of visibility $v$ is given by  $\rho_v=v\proj{\Phi^+_d}+(1-v)\I/d^2$ where $ |\Phi^+_d\rangle_{AB}
=
\frac{1}{\sqrt{d}}\sum_{i=0}^{d-1}\ket{i}_A\ket{i}_B$ and the idealised measurements are given by
\begin{eqnarray}
    Z_d=\{\proj{i}\}_{i=0}^{d-1}, \quad X_d=\{\proj{e_i}\}_{i=0}^{d-1}
\end{eqnarray}
such that $\ket{e_j}=\frac{1}{\sqrt{d}}(\sum_{k=0}^{d-1}\omega^{jk}\ket{k})$ where $\omega=e^{2\pi i/d}$.

However, in realistic conditions, the detectors are inefficient; that is, they can only detect $\eta$ fraction of the incoming subsystems. In the standard QKD protocols, one can invoke the fair-sampling assumption as the measurements are trusted. However, in the distrustful scenario, the attacker can exploit the inefficient detectors to fake a secure key between Alice and Bob \cite{DIQKDrev1,DIQKDrev2,DIQKDrev3}. Consequently, the ideal measurements of Alice for $a=0,\ldots,d-1$ are given by
\begin{eqnarray}
    M_{a|0}=\eta_A\proj{a},\quad M_{a|1}=\eta_A\proj{e_a}
\end{eqnarray}
and the additional outcomes $M_{d|x}=(1-\eta_A)\I$ corresponds to the non-detected outcome. Similarly, Bob's ideal measurements are the same as above, with a factor $\eta_B$ instead of $\eta_A$. 

Now, just like the BBM protocol, we extract the key from Alice's input $x=0$. For this purpose, let us first compute the conditional entropies $H(A_i|B_i)\geq H(A_i|B)$, as measurements increase the entropy, for $i=0,1$. One can straightaway obtain from the observed correlations $\vec{p}$ that 
\begin{equation}\label{condent}
   H(A_i|B_i)
=h_2(\eta_A)
+\eta_A(1-\eta_B)\log_2 d
-\eta_A\eta_B q_{v,d}
\end{equation}
where $h_2(x)$ is the binary entropy function and $q_{v,d}=\frac{1+(d-1)v}{d}
\log_2\!\left(\frac{1+(d-1)v}{d}\right)
+\frac{(d-1)(1-v)}{d}
\log_2\!\left(\frac{1-v}{d}\right)$. As we observe from \eqref{ratetrust}, let us compute the maximal overlap term $\tilde{\mathbbm{O}}\equiv\max_{a,a'}
\operatorname{Tr}\!\left(\tilde{M}_{a|0}\tilde{M}_{a'|1}\right)$, which is
\begin{eqnarray}
\tilde{\mathbbm{O}}
=
\begin{cases}
d(1-\eta_A)^2,
& 0\leq \eta_A\leq\dfrac{d}{d+1},\\[3mm]
\dfrac{\eta_A^2}{d},
& \dfrac{d}{d+1}\leq\eta_A\leq 1.
\end{cases}
\end{eqnarray}

Consequently, the key-rate with absolute trust in the measurements can simply be computed using $ r_{\infty,\tau=0}^{(x)}\geq-\log\tilde{\mathbbm{O}}-\sum_{i=0,1}H(A_i|B_i)$ \eqref{condent}.
To obtain the key-rate in the distrustful case \eqref{mainrate}, we compute the correction term $\Delta(\{\tilde{M}_{a|x}\},\{\tilde{M}_{a'|x'}\},\tau)$ from \eqref{correction}. We provide the analytical expressions of this term in Appendix B. 

Let us analyse the obtained key rate in the distrustful case for different levels of trust, dimensions, detection efficiencies and visibility [see Fig. \ref{fig1}]. For simplicity, we consider that Alice's and Bob's detectors have the same efficiency, $\eta_A=\eta_B=\eta$. Considering perfect detectors and sources, that is, $\eta=v=1$ or $\eta=.9$, $v=1$ [See Fig. \ref{fig1} (a) and (b)], we observe that higher dimensions allow obtaining a secure key between Alice and Bob with a lower level of trust. Moreover, we find that most current practical QKD protocols are based on the BBM (or BB84) and thus one needs to trust the devices more than $98\%$ when the detection efficiency is $\eta=.95$ and $v=1$, which is not yet attainable even in the best possible setups. Strikingly, when considering current attainable values $\eta=.85$ with $v=.99$, we find that any distrust in the measurement devices will erase the secure key. In Fig. \ref{fig1} (c) and (f), we can find the critical detection efficiencies required to set up a secure key between Alice and Bob when one trusts their devices $90\%$ and $95\%$, respectively. 
In Figure \ref{fig1} (d), considering perfect detectors, we evaluate the key rate for various dimensions and levels of trust. Finally, in \ref{fig1} (e), we analyse the effect of visibility on the key rate for various dimensions at $95\%$ trust while the detectors are considered to be perfect.

{{\it{Discussions---}} In this work, we introduced the idea of DEQKD protocols. We then found a general formula for the lower bound of all classical one-way reconciliation protocols, which can be simply expressed as the key-rate when the devices are fully trusted, with a correction term depending on the amount of trust in the devices. We finally concluded that currently deployed QKD systems are vulnerable to attacks that can not be mitigated. 
Several follow-up problems remain open in this direction. The most important of them would be to find tighter lower bounds to the key-rate formula analytically or numerically. The security of the current protocol is against collective attacks, thus, proving the security against coherent attacks using generalised entropy accumulation theorem is important. The next problem would be to find the formula for the upper bound to the key-rate in the DEQKD scenario, as this would quantify the least amount of trust needed to do practical QKD. Applying this formulation to other protocols such as MDI-QKD \cite{mdi2012, mdi2013} or repeater-based QKD \cite{repeater1998,repeater2003,repeater} would be interesting.

%

\onecolumngrid

\appendix

\section{General formula for the lower bound on the key-rate in the distrustful DEQKD scenario}

Consider the standard QKD scenario with one-way classical reconciliation such that the measurements are trusted up to some parameter $1-\tau$, where $\tau$ is the distrust parameter. This implies that for all $a,x$
\begin{equation}\label{eq1}
    \dl{M_{a|x}-\tilde{M}_{a|x}}_2\leq \tau
\end{equation}
where $\tilde{M}_{a|x}$ is the ideal measurement POVM while $M_{a|x}$ is the actual measurement POVM in the setup. 

It is well-known that the asymptotic key rate obtainable in this setup is lower-bounded using the Devetak-Winter (DW) rate as
    \begin{equation}\label{rate}
         r_{\infty,\tau}^{(x)}\geq H(A_x|E)-H(A_x|B)
    \end{equation}
    where $r_{\infty,\tau}^{(x)}$ denotes the rate when the key is extracted from the input $x$. Let us now consider the entropic uncertainty relation for arbitrary measurements among three parties $A,B,E$ introduced in \cite{Coles_2014}, given as

\begin{eqnarray}
    H(A_{x'}|B)+H(A_{x}|E)\geq q(\rho_A),
\end{eqnarray}
where $ q(\rho_A)=\max\left\{q(\rho_A,A_{x'},A_{x}),q(\rho_A,A_{x},A_{x'})\right\},$
with $ q(\rho_A,A_{x'},A_{x})=-\sum_{a'} p_A(a'|x')\log_2 h_{a'}(A_{x'},A_{x})$, and $q(\rho_A,A_{x},A_{x'})=-\sum_{a} p_A(a|x)\log_2 h_a(A_{x},A_{x'})$. Here, $h_{a'}(A_{x'},A_{x})=\left\|\sum_k M_{a|x}M_{a'|x'}M_{a|x}\right\|_\infty$, and $p_A(k|x)=\operatorname{Tr}(\rho_A M_{k|x}).$ However, as shown in \cite{Coles_2014}, the above relation can be further simplified to
\begin{eqnarray}
    H(A_{x'}|B)+H(A_{x}|E)\geq -\log \Big( \max_{a,a'}c_{a,a'}\Big): \qquad c_{a,a'}=\dl{\sqrt{M_{a|x}}\sqrt{M_{a'|x'}}}^2_{\infty}.
\end{eqnarray}
Using the fact that $\dl{T}_{\infty}\leq\dl{T}_2$ for any matrix $T$, we obtain from \eqref{rate} that
\begin{eqnarray}
    r_{\infty,\tau}^{(x)}\geq -\log \Big( \max_{a,a'}\Tr(M_{a|x}M_{a'|x'})\Big)-H(A_x|B)- H(A_{x'}|B).
\end{eqnarray}

\begin{thm*}\label{thm1}
    Consider the pair of measurements corresponding to inputs $x,x'$. The maximum overlap of the measurement elements is given by 
    \begin{eqnarray}\label{maineq1}
        \max_{a,a'}\Tr(M_{a|x}M_{a'|x'})\leq  \max_{a,a'}\left[\Tr(\tilde{M}_{a|x}\tilde{M}_{a'|x'})+3\tau\left(\dl{\tilde{M}_{a|x}}+\dl{\tilde{M}_{a'|x'}}\right)\right]+6\tau^2\nonumber\\
        \leq  \max_{a,a'}\Tr(\tilde{M}_{a|x}\tilde{M}_{a'|x'})+3\tau\left(\max_{a}\dl{\tilde{M}_{a|x}}+\max_{a'}\dl{\tilde{M}_{a'|x'}}\right)+6\tau^2
    \end{eqnarray}
    where $\tilde{M}_{a|x},\tilde{M}_{a'|x'}$ are the ideal measurement elements.
\end{thm*}
\begin{proof}
    Considering the relation \eqref{eq1}, we obtain that
    \begin{equation}
         \dl{M_{a|x}-\tilde{M}_{a|x}}+ \dl{M_{a'|x'}-\tilde{M}_{a'|x'}}\leq 2\tau
    \end{equation}
    which using triangle inequality implies that
    \begin{equation}
       \dl{\tilde{M}_{a|x}-\tilde{M}_{a'|x'}}\leq 2\tau+ \dl{M_{a|x}-M_{a'|x'}}.
    \end{equation}
Squaring both sides, we obtain
 \begin{equation}
\Tr(\tilde{M}_{a|x}^2)+\Tr(\tilde{M}_{a'|x'}^2)-2\Tr(\tilde{M}_{a|x}\tilde{M}_{a'|x'})\leq \Tr(M_{a|x}^2)+\Tr(M_{a'|x'}^2)-2\Tr(M_{a|x}M_{a'|x'})+4\tau\dl{M_{a|x}-M_{a'|x'}}+4\tau^2    \end{equation}
which, on rearranging the terms, gives us
\begin{equation}\label{eq2}
        2\Tr(M_{a|x}M_{a'|x'})\leq    \Tr(M_{a|x}^2)+\Tr(M_{a'|x'}^2)-\Tr(\tilde{M}_{a|x}^2)-\Tr(\tilde{M}_{a'|x'}^2)+4\tau^2+2\Tr(\tilde{M}_{a|x}\tilde{M}_{a'|x'})+4\tau\dl{M_{a|x}-M_{a'|x'}}.
    \end{equation}
    Let us now bound the quantity $\Tr(M_{a|x}^2)$ (similarly $\Tr(M_{a'|x'}^2)$) . For this purpose, we first observe that
    \begin{equation}\label{eq3}
        \dl{M_{a|x}}\leq \dl{\tilde{M}_{a|x}}+ \dl{M_{a|x}-\tilde{M}_{a|x}}\leq  \dl{\tilde{M}_{a|x}}+\tau.
    \end{equation}
    Consequently, we obtain from \eqref{eq2} that
    \begin{equation}
        2\Tr(M_{a|x}M_{a'|x'})\leq 2\tau\dl{\tilde{M}_{a|x}}+2\tau\dl{\tilde{M}_{a'|x'}}   +4\tau^2+2\Tr(\tilde{M}_{a|x}\tilde{M}_{a'|x'})+4\tau\dl{M_{a|x}-M_{a'|x'}}.
    \end{equation}
    Again using \eqref{eq3}, we can bound $\dl{M_{a|x}-M_{a'|x'}}$, such that the above expression can be simplified to
    \begin{eqnarray}
         2\Tr(M_{a|x}M_{a'|x'})\leq 2\tau\dl{\tilde{M}_{a|x}}+2\tau\dl{\tilde{M}_{a'|x'}}   +4\tau^2+2\Tr(\tilde{M}_{a|x}\tilde{M}_{a'|x'})+4\tau\left(\dl{\tilde{M}_{a|x}}+\dl{\tilde{M}_{a'|x'}}+2\tau\right).
    \end{eqnarray}
    This finally gives us
    \begin{equation}
        \Tr(M_{a|x}M_{a'|x'})\leq   \Tr(\tilde{M}_{a|x}\tilde{M}_{a'|x'})+3\tau\left(\dl{\tilde{M}_{a|x}}+\dl{\tilde{M}_{a'|x'}}\right) +6\tau^2.
    \end{equation}
    which using the fact that $\max\{A+B\}\leq \max A+\max B$ gives us \eqref{maineq1}.
\end{proof}

\section{Distrustful generalised BBM protocol}

The state shared between Alice and Bob of visibility $v$ is given by  $\rho_v=v\proj{\Phi^+_d}+(1-v)\I/d$ where $ |\Phi^+_d\rangle_{AB}
=
\frac{1}{\sqrt{d}}\sum_{i=0}^{d-1}\ket{i}_A\ket{i}_B$ and the idealised measurements are given by
\begin{eqnarray}
    M_{a|0}=\eta_A\proj{a},\quad M_{a|1}=\eta_A\proj{e_a},\quad a=0,1,\ldots,d-1\quad|\quad M_{d|x}=(1-\eta_A)\I\quad(x=0,1)
\end{eqnarray}
where $\ket{e_j}=\frac{1}{\sqrt{d}}(\sum_{k=0}^{d-1}\omega^{jk}\ket{k})$ such that $\omega=e^{2\pi i/d}$.
Similarly, Bob's ideal measurements are given by
\begin{eqnarray}
    M_{b|0}=\eta_B\proj{b},\quad M_{b|1}=\eta_B\proj{e_b},\quad a=0,1,\ldots,d-1\quad|\quad M_{b|y}=(1-\eta_B)\I\quad(y=0,1).
\end{eqnarray}
Alike the BBM protocol, we extract the key from Alice's input $x=0$. Let us compute the  overlap term $-\log \Big( \max_{a,a'}\Tr(M_{a|0}M_{a'|1})\Big)$. For this purpose, we first observe that
\begin{eqnarray}
    \Tr\!\left(M_{a|0}M_{a'|1}\right)
=
\begin{cases}
\dfrac{\eta_A^2}{d},
& a,a'=0,\ldots,d-1,\\[2mm]
\eta_A(1-\eta_A),
& a=0,\ldots,d-1,\;a'=d,\\[2mm]
\eta_A(1-\eta_A),
& a=d,\;a'=0,\ldots,d-1,\\[2mm]
d(1-\eta_A)^2,
& a=a'=d.
\end{cases}
\end{eqnarray}
Consequently, we have that
\[
\max_{a,a'}
\operatorname{Tr}\!\left(\tilde{M}_{a|0}\tilde{M}_{a'|1}\right)
=
\begin{cases}
d(1-\eta_A)^2,
& 0\leq \eta_A\leq\dfrac{d}{d+1},\\[3mm]
\dfrac{\eta_A^2}{d},
& \dfrac{d}{d+1}\leq\eta_A\leq 1.
\end{cases}
\]
Consequently, the key-rate with absolute trust in the measurements is given by
\begin{eqnarray}\label{ratet=0}
    r^{0}_{\infty,\tau=0}
\geq
\begin{cases}
-\log d-2\log(1-\eta_A)-\sum_{x=0,1}H(A_x|B_x),
& 0\leq \eta_A\leq\dfrac{d}{d+1},\\[3mm]
\log d-2\log\eta_A-\sum_{x=0,1}H(A_x|B_x),
& \dfrac{d}{d+1}\leq\eta_A\leq 1
\end{cases}
\end{eqnarray}
where $H(A_0|B_0)=H(A_{1}|B_{1})$ for the above considered scenario given in \eqref{condent}. 
To obtain the key-rate in the distrustful case, let us compute the correction term $\Delta(\{\tilde{M}_{a|x}\},\{\tilde{M}_{a'|x'}\},\tau)=\left(3\tau\left(\max_{a}\dl{\tilde{M}_{a|x}}+\max_{a'}\dl{\tilde{M}_{a'|x'}}\right)+6\tau^2\right)/\max_{a,a'}\Tr(\tilde{M}_{a|x}\tilde{M}_{a'|x'})$ from \eqref{correction},
\begin{eqnarray}\label{corr}
     \Delta(\{\tilde{M}_{a|x}\},\{\tilde{M}_{a'|x'}\},\tau)=
     \begin{cases}
\frac{6\tau}{\sqrt{d}(1-\eta_A)}+\frac{6\tau^2}{d(1-\eta_A)^2},
& 0\leq \eta_A<\dfrac{\sqrt{d}}{\sqrt{d}+1},\\[3mm]
\frac{6\tau \eta_A+6\tau^2}{d(1-\eta)^2},
&\dfrac{\sqrt{d}}{\sqrt{d}+1} \leq\eta_A< \dfrac{d}{d+1},\\[3mm]
6d\tau\left(1+\frac{\tau}{\eta_A}\right),
& \dfrac{d}{d+1}\leq\eta_A\leq 1.
\end{cases}
\end{eqnarray}
Thus, Eq. \eqref{ratet=0} and Eq. \eqref{corr} allow us to obtain the key rate plots in Fig \ref{fig1} of the manuscript.

\end{document}